\documentclass[a4paper,12pt]{article}
\usepackage{jheppub_modified_green}%
\usepackage{graphicx}
\usepackage{epsfig}
\usepackage{amsfonts}
\usepackage{amsmath}
\usepackage{amssymb}
\usepackage{enumitem}
\usepackage{bbold}
\usepackage{amsthm}
\usepackage{subfig}
\usepackage{bm}
\usepackage{hyperref}
\usepackage{mathrsfs}
\usepackage{bbm}
\usepackage{cancel}
\usepackage{xcolor}
\usepackage{accents}
\usepackage{relsize}
\usepackage{float, slashed, graphicx, amssymb, amsmath}
\usepackage{shuffle}
\usepackage{tikz}
\usetikzlibrary{hobby,calc,shapes, positioning, backgrounds,trees, decorations, decorations.markings, decorations.pathmorphing, arrows, shapes.geometric, snakes}
\usepackage{tikz-feynman}
\tikzfeynmanset{compat=1.1.0}
\tikzfeynmanset{
graviton/.style={circle, draw=green!60, fill=green!5, very thick, minimum size=7mm}
}
\tikzfeynmanset{
codot/.style={/tikz/shape=circle,
/tikz/fill=red,/tikz/minimum size=0.1cm,/tikz/inner sep=1.8pt}
}
\tikzfeynmanset{sb/.style=
{/tikz/shape=circle,
/tikz/fill=black,
 /tikz/minimum size=0.3cm,/tikz/inner sep=1.pt
 } }
 \tikzfeynmanset{rsb/.style=
{/tikz/shape=circle,
/tikz/fill=red,
 /tikz/minimum size=0.3cm,/tikz/inner sep=1.pt
 } }
\tikzfeynmanset{myblob/.style=
{/tikz/shape=ellipse,
/tikz/fill=red,
 /tikz/minimum width=0.5cm,
 } }
 \tikzfeynmanset{myblobFF/.style=
{/tikz/shape=circle,
/tikz/fill=black,
 /tikz/minimum width=0.25cm,
 } }
 \tikzfeynmanset{myblob2/.style=
{/tikz/shape=rectangle,
/tikz/fill=black,
 /tikz/minimum width=0.1cm,/tikz/inner sep=1.8pt
 } }
  \tikzfeynmanset{fp/.style=
{/tikz/shape=rectangle,
/tikz/fill=red,
 /tikz/minimum width=0.1cm,/tikz/inner sep=1.8pt
 } }
  \tikzfeynmanset{myvertex/.style=
{/tikz/shape=circle,
/tikz/fill=black,
 /tikz/minimum width=0.05cm,
 } }
\tikzset{box/.pic={\filldraw[fill=black]  (0,0) circle (2.5pt);
				   \filldraw [fill=black] (0.5,0) circle (2.5pt);
			       \draw [line width=5pt] (0,0) -- (0.5,0);}}

\tikzset{wiggle/.style={decorate, decoration=snake}}

 \tikzfeynmanset{HV/.style=
{/tikz/shape=circle,
/tikz/fill={rgb:black,1;white,2},
 /tikz/minimum size=0.01cm,/tikz/inner sep=1.8pt
 } }

\usepackage[T1]{fontenc} 

\def\sc#1{\overline{#1}}
\makeatletter
\newcommand \UPlus {\mathop {\operator@font \uplus }\limits }
\makeatother
\makeatletter
\newcommand \Bigcup {\mathop {\operator@font \bigcup }\limits }
\makeatother
  \def\LabelNote#1{}
 \def\Label#1{\label{#1}%
  \smash{\hbox to\phipt{\raise1ex\hbox{\tiny[#1]}\hss}}}
  
  \def\Cdot{{\cdot}}
  
\def\nn{\nonumber}

\newcommand{\blue}{\color{blue}}

\newcommand{\white}{\color{white}}

\def\spa#1.#2{\left\langle\!\langle#1\,#2\right\rangle\!\rangle}
\def\spb#1.#2{\left[#1\,#2\right]}

\def\be{\begin{equation}}
\def\ee{\end{equation}}
\def\bea{\begin{eqnarray}}
\def\eea{\end{eqnarray}}  
\allowdisplaybreaks

\newcommand{\npre}{\mathcal{N}}  

\newcommand{\commut}{\Gamma}

\newcommand{\la}{\langle\!\langle}
\newcommand{\ra}{\rangle\!\rangle}
\newcommand{\mdot}{{\cdot}}
\newcommand{\veps}{\varepsilon}
\makeatletter
\newcommand*{\bigcdot}{}
\DeclareRobustCommand*{\bigcdot}{%
  \mathbin{\mathpalette\bigcdot@{}}%
}
\newcommand*{\bigcdot@scalefactor}{.6}
\newcommand*{\bigcdot@widthfactor}{1.25}
\newcommand*{\bigcdot@}[2]{%
  \sbox0{$#1\vcenter{}$}
  \sbox2{$#1\cdot\m@th$}%
  \hbox to \bigcdot@widthfactor\wd2{%
    \hfil
    \raise\ht0\hbox{%
      \scalebox{\bigcdot@scalefactor}{%
        \lower\ht0\hbox{$#1\bullet\m@th$}%
      }%
    }%
    \hfil
  }%
}
\makeatother
\newcommand{\dd}{\bigcdot}

\begin{document} 
 
\title{Color-Kinematic Numerators for Fermion Compton Amplitudes}

\author{N. Emil J. Bjerrum-Bohr$\mbox{}^{a}$,}
\author{Gang Chen$\mbox{}^{a}$,}
\author{Yuchan Miao$\mbox{}^{a}$,}
\author{Marcos Skowronek$\mbox{}^{b}$}
\affiliation{$\mbox{}^{a}$Niels Bohr International Academy,
Niels Bohr Institute, University of Copenhagen,\\
Blegdamsvej 17, DK-2100 Copenhagen \O, Denmark}
\affiliation{$\mbox{}^{b}$Department of Physics, Brown University, Providence, RI 02912, USA}
\emailAdd{bjbohr@nbi.dk}
\emailAdd{gang.chen@nbi.ku.dk}
\emailAdd{sxh962@alumni.ku.dk}
\emailAdd{marcos$\underline \ $skowronek$\underline \ $santos@brown.edu}

\abstract{
We introduce a novel approach to compute Compton amplitudes involving a fermion pair inspired by Hopf algebra amplitude constructions. This approach features a recursive relation employing quasi-shuffle sets, directly verifiable by massive factorization properties. We derive results for minimal gauge invariant color-kinematic numerators with physical massive poles using this method. We have also deduced a graphical method for deriving numerators that simplifies the numerator generation and eliminates redundancies, thus providing several computational advantages.
} 


\maketitle\tableofcontents

\section{Introduction}\medskip
The Kawai-Lewellen-Tye string theory relations \cite{Kawai:1985xq} combine gravity and Yang-Mills theories \cite{Bern:1998sv} and have led to the important discovery of a color and kinematics duality \cite{Bern:2008qj,Bjerrum-Bohr:2009ulz,Bern:2010ue}; a cornerstone for efficient computation of amplitudes and motivation for examining core principles of quantum particle scattering \cite{Bern:2019prr}.  

A considerable attraction over a decade has been finding a kinematic algebra that mimics that of the color group \cite{Bern:2011ia,Monteiro:2011pc,Bjerrum-Bohr:2012kaa}, and recently, there has been a breakthrough with the systematic framework developed by refs. \cite{Chen:2019ywi, Chen:2021chy, Brandhuber:2021kpo}. This framework uses infinite-dimensional combinatorics \cite{Brandhuber:2021bsf, Brandhuber:2022enp} and a direct numerator construction approach connected to quasi-shuffle Hopf algebras. The numerator technology developed this way is universal across many theories, including those with a spinless massive field, such as the heavy-mass effective field theory refs. \cite{Georgi:1990um, Luke:1992cs, Neubert:1993mb, Manohar:2000dt, Damgaard:2019lfh,Brandhuber:2021kpo, Brandhuber:2021eyq}, scalar Yang-Mills theory with finite mass \cite{Chen:2022nei, Cao:2022vou}, the $\alpha'F^3+\alpha'^2F^4$ higher-derivative interaction theory \cite{Chen:2023ekh} (see also refs. \cite{Garozzo:2018uzj, Chen:2022shl,Pavao:2022kog,Carrasco:2022sck,Chen:2023dcx,Li:2023wdm,Bonnefoy:2023imz,Brown:2023srz,Carrasco:2021ptp,Carrasco:2022jxn,Garozzo:2024myw}), and the $\rm DF^2+YM$ theory \cite{Chen:2024gkj} (see as well refs. \cite{Huang:2016tag,Johansson:2017srf,Azevedo:2018dgo}).

This paper studies the kinematic algebra for the gauge invariant color-kinematic numerator associated with the generalized massive Compton amplitude with a fermion pair. Although this amplitude is considerably more complex than the spinless massive field case (see also refs. \cite{Bjerrum-Bohr:2019nws,Bjerrum-Bohr:2020syg,Edison:2020ehu,Lin:2022jrp}), we show the realization of the algebra as a closed-form quasi-shuffle product. The amplitudes discussed are of theoretical interest as they allow for structural insight into fermion factorization limits and are useful when generating inspiration for bootstrap methods for general spin amplitudes \cite{Bjerrum-Bohr:2023jau,Bjerrum-Bohr:2023iey}. We thus hope that progress in this area could potentially help develop computational methods for spinning black hole merger observables, for instance, using the modern amplitude techniques pioneered in refs. such as \cite{Bjerrum-Bohr:2013bxa,Bjerrum-Bohr:2018xdl,Cheung:2018wkq,Bern:2019nnu,Bjerrum-Bohr:2021din,Brandhuber:2021eyq, Bjerrum-Bohr:2021wwt,Bjerrum-Bohr:2022ows}.

We present our work in this paper as follows. First, we introduce the kinematic Hopf algebra in section \ref{sec:algebra} and define a new evaluation map for massive Compton amplitudes with fermion pairs. Next, we demonstrate the validity of this formalism through computation and employ factorization properties to outline a proof of the construction.
In section \ref{sec:map2}, we provide an evaluation map that simplifies numerators using the on-shell condition of the external massive fermion states. This graphical method streamlines numerator generation and thereby delivers computational benefits.
Finally, in the final section, we provide an outlook for future research.
\section{Color-kinematic fermion pair numerators}\label{sec:algebra}
This section will deduce and demonstrate our algebraic construction for numerators. We will start by setting up a convenient and compact notation. A key element is an $n$-point massive fermion amplitude representation in terms of minimal color-kinematic numerators with $(n\!-\!3)!$ elements involving nested commutators. As an illustration of this, we consider ordered cubic vertex diagrams associated with gluon lines, $1, 2,$ and $3$ connected through a propagator to gluon legs $4,5,\ldots,n\!-\!2$ and two massive fermions $\sc{n{-}1},\sc n$.
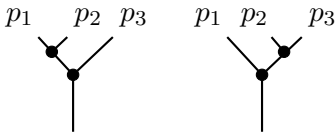
\begin{equation}
\begin{aligned}
    \begin{tikzpicture}[baseline={([yshift=-0.8ex]current bounding box.center)}]\tikzstyle{every node}=[font=\small]    
   \begin{feynman}
    \vertex (a)[]{$~$};
      \vertex [above=0.9cm of a](b)[dot]{};
     \vertex [left=0.7cm of b](c);
     \vertex [left=0.28cm of b](c23);
     \vertex [above=0.23cm of c23](v23)[dot]{};
    \vertex [above=.5cm of c](j1){$p_1$};
    \vertex [right=.9cm of j1](j2){$p_2$};
    \vertex [right=0.6cm of j2](j3){$p_3$};
   	 \diagram*{(a) -- [thick] (b),(b) -- [thick] (j1),(v23) -- [thick] (j2),(b)--[thick](j3)};
    \end{feynman}  
  \end{tikzpicture} &&  \begin{tikzpicture}[baseline={([yshift=-0.8ex]current bounding box.center)}]\tikzstyle{every node}=[font=\small]    
   \begin{feynman}
    \vertex (a)[]{$~$};
      \vertex [above=0.9cm of a](b)[dot]{};
     \vertex [left=0.7cm of b](c);
     \vertex [right=0.29cm of b](c23);
     \vertex [above=0.23cm of c23](v23)[dot]{};
    \vertex [above=.5cm of c](j1){$p_1$};
    \vertex [right=.6cm of j1](j2){$p_2$};
    \vertex [right=0.9cm of j2](j3){$p_3$};
   	 \diagram*{(a) -- [thick] (b),(b) -- [thick] (j1),(v23) -- [thick] (j2),(b)--[thick](j3)};
    \end{feynman}  
  \end{tikzpicture}  \nn \\ 
\end{aligned} 
\end{equation}
The two possible propagators respectively are $p_{12}^2p_{123}^2$ and $p_{23}^2p_{123}^2$, (defining $p_{ij} \equiv p_i+p_j$ and $p_{ij\ldots k} \equiv p_i+p_j+...+p_k$) and it is easy to see that the color-kinematic duality holds writing out the Jacobi identity for numerators. %
\begin{align}
\begin{split}
\label{eq:newDCg}
A_{\!f}(1,2,3,\sc{n{-}1},\sc n)&\! =\! {\npre_{\!f}([[1,2],3],\sc{n{-}1},\sc n)\over p_{12}^2p_{123}^2}\!+\!  {\npre_{\!f}([1,[2,3]],\sc{n{-}1},\sc n)\over p_{23}^2p_{123}^2}\!
=\! \sum_{\commut \in \rho(3)} \!\!{\npre_{\!f}(\commut,\sc{n{-}1},\sc n)\over d_\commut}\!\,,\end{split}
\end{align} 
where $\Gamma\in
\rho(3)\equiv\{[[1,2],3],[1,[2,3]]\}$ and $d_\commut$ are massless propagators associated with numerators. This notation readily extends to any number of legs as follows
\begin{align}
\begin{split}
\label{eq:newDCg}
	A_{\!f}(1\ldots n{-}2,\sc{n{-}1},\sc n)&\, =\, \sum_{\commut \in \rho(n\!-\!2)} {\npre_{\!f}(\commut,\sc{n{-}1},\sc n)\over d_\commut}\,,
\end{split}
\end{align}
\\[10pt]
where now $\Gamma$ sums over the possible $(n\!-\!3)!$ configurations of ordered nested numerators associated with the $n\!-\!2$ gluon legs.
We can also compute graviton amplitudes using the double-copy. In this case, we write the $n$ point amplitude as
\begin{align}
\begin{split}
\label{eq:newDCgr}
	M_{\!f}(1\ldots n{-}2,\sc{n{-}1},\sc n)&\, =\, \sum_{\commut\in \tilde{\rho}(n\!-\!2)} {\npre_{\!f}(\commut,\sc{n{-}1},\sc n)\npre(\commut,\sc n,\sc{n{-}1}) \over d_\commut}\, .
\end{split}
\end{align}
We denote sums over configurations by $\tilde{\rho}(n\!-\!2)$, where the `tilde' indicates that the sum includes all possible nested unordered commutator configurations associated with gravitons labels $\{ 1, \ldots \, n{-}2 \}$.

Using the gravity momentum ${\cal S}$-kernel \cite{Bjerrum-Bohr:2010diw,Bjerrum-Bohr:2010pnr,Bjerrum-Bohr:2010mia,Bjerrum-Bohr:2010kyi,Bjerrum-Bohr:2010mtb}, we know that we can always generate generic color-kinematic numerators using the following prescription  \cite{Bjerrum-Bohr:2010pnr,Bjerrum-Bohr:2012kaa}
\begin{align}\label{eq:KLT}
\!\!\!\!   \npre_{\!f}([[[[\cdots[1,2],3],\ldots], n\!-\!2],\sc{n\!-\!1},\sc{n})\!=\! \sum_{\beta\in S_{n\!-\!3}} \mathcal{S}[1\ldots n\!-\!2 |1\beta ]A(1\beta,\sc{n\!-\!1},\sc{n}) \,.
\end{align}
In the sum, $S_{n\!-\!3}$ denotes permutations of gluon indices $\{2,3,\ldots, n\!-\!2\}$ and we work with the momentum ${\cal S}$-kernel definition employed in \cite{Brandhuber:2021bsf}. We will, for compactness in the following, denote the left-nested commutator of gluons as $[12\ldots n\!-\!2]$, {\it {\it i.e.}}, $[1234]\equiv[[[1,2],3],4]$. (For more details, see also refs. \cite{Carrasco:2016ldy,Frost:2020eoa}). When considering factorization, the following recursive property of the gravity momentum $\cal S$-kernel is ideal 
\begin{align}\label{Srec}
    \mathcal{S}[1\ldots j|1\beta_L j\beta_R]=2(p_{\Theta_L(j)}\Cdot p_j)\, \mathcal{S}[1\ldots j{-}1|1\beta_L \beta_R] \, .
\end{align}
In this context, the symbol $\Theta_L(j)$ denotes the indices that come before the index $j$ in the set $\{1, \beta_{L}\}$, with an index order that is lower than that of $j$. With these conventions, the expression $\mathcal{S}[1\alpha |1\beta]$ only depends on the massless momentum and cancels out all the massless propagators in the amplitude, although not manifestly. 

\subsection{Numerator examples at four and five points}
Before diving into the kinematic algebra, we will explore a manifestly gauge-invariant double-copy numerator form of the quark-gluon Compton amplitude derived from Feynman rules. We have adopted the following convention for the three-point amplitudes
\begin{equation}
\begin{aligned}
     \begin{tikzpicture}[baseline=(current bounding box.center)]
  \begin{feynman}
       \vertex (a){\(\overline 3\)};
       \vertex [right=2.cm of a](b){\(\overline 2\)};
       \vertex[right=1.0cm of a](c);
       \vertex[above=1.0cm of c](d){\(1\)};
       \diagram* { 
                 (b) -- [fermion] (c) -- [fermion] (a),
                 (c) -- [gluon] (d)
                };
  \end{feynman}
\end{tikzpicture} 
= \frac{1}{2}{\bar v \slashed{\varepsilon}_{1} u} \, , &&
\begin{tikzpicture}[scale=0.8, transform shape, baseline=(current bounding box.center)]
\begin{feynman}
  \vertex (c0);
  \vertex [above=1.2cm of c0] (a) {\(1\)}; 
  \vertex [below left=0.55cm and 1.0cm of c0] (b) {\(3\)};
  \vertex [below right=0.55cm and 1.0cm of c0] (c) {\(2\)}; 
  \diagram* {
    (c0) -- [gluon] (a),
    (c0) -- [gluon] (b),
    (c0) -- [gluon] (c)
  };
\end{feynman}
\end{tikzpicture}
{\begin{minipage}{0.3\textwidth}
            \begin{align}
              & =(\varepsilon_{1} \mdot \varepsilon_{3})(p_1 \mdot \varepsilon_{2})+(\varepsilon_{2} \mdot \varepsilon_{3})(p_3 \mdot \varepsilon_{1})\nn\\
&~~~~~+(\varepsilon_{1} \mdot \varepsilon_{2})(p_2 \mdot \varepsilon_{3})\,,\nn
            \end{align}
        \end{minipage}}
\end{aligned}
\end{equation}
where $\slashed \veps\equiv \gamma^\mu \veps_\mu$, for a given gluon polarization vector, $\bar v$, and $u$ denote  the incoming and outgoing  massive fermions states with on-shell conditions $
    \bar{v}(\slashed{p}_3-m)=0 \, , 
    (\slashed{p}_{2}+m) u=0, \, {\rm and\ mass\ }m.
$
The external momenta are considered outgoing. 
Our focus is on the first non-trivial case, the four-point quark-gluon amplitude, which is featured by two contributing color-ordered Feynman diagrams:
\begin{align}
    & \begin{tikzpicture}
     [baseline={([yshift=-0.4ex]current bounding box.center)}]\tikzstyle{every node}=[font=\small] 
    \begin{feynman}
        \vertex (4) {\(\sc 4\)};
        \vertex[right=1cm of 4] (14);
        \vertex[right=1cm of 14] (23);
        \vertex[right=1cm of 23] (3){\(\overline 3\)};
        \vertex[above=1cm of 14] (1) {\(1\)};
        \vertex[above=1cm of 23] (2) {\(2\)};
            \diagram* {
                {[edges=fermion]
                 (3) -- (23) -- (14) -- (4)
                },
                    (23) -- [gluon] (2),
                    (14) -- [gluon] (1),
                        };
    \end{feynman}
    \end{tikzpicture}
    ={1\over 4 (p_{14}^2-m^2)}
    (\bar{v}\slashed{\varepsilon}_{1} (\slashed{p}_{14}+m) \slashed{\varepsilon}_{2}u) \, ,
    \\
    & \begin{tikzpicture}
    [baseline={([yshift=-2.1ex]current bounding box.center)}]\tikzstyle{every node}=[font=\small] 
    \begin{feynman}
        \vertex (4) {\(\sc 4\)};
        \vertex[right=1.5cm of 4] (1234);
        \vertex[right=1.5cm of 1234] (3){\(\overline 3\)};
        \vertex[above=1cm of 1234] (12);
        \vertex[above right=1cm of 12] (2){\(2\)};
        \vertex[above left=1cm of 12] (1){\(1\)};
 \diagram* {
 {
 [edges=fermion]
 (3) -- (1234) -- (4)
 },
(1234) -- [gluon] (12),
 (12) -- [gluon] (1),
 (12) -- [gluon] (2),
        };
    \end{feynman}
    \end{tikzpicture} 
     \begin{aligned}
    &=
-\frac{1}{2p_{12}^2}
\Big[\left(p_{2} \mdot \varepsilon_{1}\right)\left(\bar{v} \slashed{\varepsilon}_{2} u\right)-\left(p_{1} \mdot \varepsilon_{2}\right)\left(\bar{v} \slashed{\varepsilon}_{1} u\right)\nn\\
&~~~~~~~~~~~+\frac{1}{2}\left(\varepsilon_{1} \mdot \varepsilon_{2}\right)(\bar{v}(\slashed{p}_{1}-\slashed{p}_{2}) u)\Big] \, .
\end{aligned}
\end{align}
The four-point KLT momentum kernel matrix as indicated by  eq.~\eqref{Srec} provides a double-copy numerator of the form (we use $\mathcal{S}[12|12]= 
2(p_{1}\mdot p_{2})$)
\begin{align}
\mathcal{N}_f([1 2], \overline 3, \overline 4)  &= 
2(p_{1}\mdot p_{2}) A_f(1 2, \overline 3, \overline 4)=
    \frac{1}{2(p_{14}^2-m^2)}
    \bar{v} 
   \Big(
   p_1 \mdot p_2\slashed{\varepsilon}_{1}(\slashed{p}_{14}+m)\slashed{\varepsilon}_2 \nn\\
    &-(p_{1}\mdot p_4)
    \left[
    2\left(p_{2} \mdot \varepsilon_{1} \right)\slashed{\varepsilon}_2-2\left(p_{1} \mdot \varepsilon_{2} \right)\slashed{\varepsilon}_1+\left(\varepsilon_{1} \mdot \varepsilon_{2}\right)(\slashed{p}_1-\slashed{p}_2)
    \right]
    \Big) u \ .
\end{align}
After simplification, this leads to a compact formulation of the double-copy numerator, expressed in terms of the field strength tensor:
\begin{equation}
    \label{4ptdouble-copy}
    \mathcal{N}_f([12], \overline 3, \overline 4)=-{1\over p_{41}^2-m^2} \Big(
   \bar{v} \dd  {(p_4\mdot F_{12})} \dd  u+{1\over 4}\bar{v} \dd F_1 \dd (p_1 \mdot F_{2})\dd  u\Big) \, .
\end{equation}
In this context, we use $F_{i}^{\mu \nu}$ to denote the Abelian field strength tensor, so that $F_{i}^{\mu \nu}\equiv p_{i}^{\mu} \varepsilon_{i}^{\nu}-\varepsilon_{i}^{\mu} p_{i}^{\nu}$. We represent the Dirac matrix product with a bullet symbol $\dd$, and we employ the simplified notation as follows:
\begin{align}
    \bar v\dd X\dd (p_1\cdot F_2)\dd Y\dd u &\equiv (\bar v\dd X\dd \gamma_\mu\dd Y \dd u)\, (p_1\cdot F_2)^{\mu},\nn\\
    \bar v\dd X\dd F_{i_1\ldots i_r}\dd Y\dd u &\equiv  (\bar v\dd X\dd \gamma_{\mu}\dd \gamma_{\nu}\dd Y\dd u)\, F_{i_1\ldots i_r}^{\mu\nu} \, .
\end{align}
Here $F_{i_1i_2\ldots i_r}^{\mu\nu}\equiv (F_{i_1}\cdot F_{i_2} \cdots F_{i_r})^{\mu\nu}$ denotes the contraction of a string of multiple field strength tensors. 

Next, we turn our attention to the five-point quark-gluon amplitude, which we can compute from the four diagram topologies:
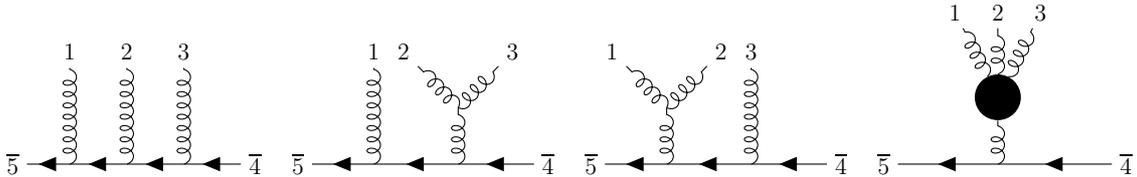
\begin{equation*}
\begin{tikzpicture}[scale=0.75, transform shape, baseline=(1.base)]
\begin{feynman}
    \vertex (1) {\(\overline 5\)};
    \vertex[right=1cm of 1] (12);
    \vertex[right=1cm of 12] (123);
    \vertex[right=1cm of 123] (1234);
    \vertex[right=1cm of 1234] (5){\(\overline{4}\)};
    \vertex[above=1.7cm of 12] (2) {\(1\)};
    \vertex[above=1.7cm of 123] (3) {\(2\)};
    \vertex[above=1.7cm of 1234] (4) {\(3\)};
    \diagram* {
        {[edges=fermion]
            (5)--(1234)--(123)--(12)--(1),
        },
        (12) -- [gluon] (2),
        (123) -- [gluon] (3),
        (1234) -- [gluon] (4),
    };
\end{feynman}
\end{tikzpicture}%
\hspace{2mm}
\begin{tikzpicture}[scale=0.75, transform shape, baseline=(1.base)]
\begin{feynman}
    \vertex (1) {\(\overline 5\)};
    \vertex[right=1.33cm of 1] (12);
    \vertex[right=1.48cm of 12] (123);
    \vertex[right=1.33cm of 123] (5){\(\overline{4}\)};
    \vertex[above=1.7cm of 12] (2) {\(1\)};
    \vertex[above=1cm of 123] (34);
    \vertex[above right=1cm of 34] (4) {\(3\)};
    \vertex[above left=1cm of 34] (3) {\(2\)};
    \diagram* {
        {[edges=fermion]
            (5)--(123)--(12)--(1),
        },
        (12) -- [gluon] (2),
        (123) -- [gluon] (34),
        (34) -- [gluon] (3),
        (34) -- [gluon] (4),
    };
\end{feynman}
\end{tikzpicture}%
\hspace{2mm}
\begin{tikzpicture}[scale=0.75, transform shape, baseline=(a.base)]
\begin{feynman}
    \vertex (a) {\(\overline 5\)};
    \vertex[right=1.33cm of a] (ab);
    \vertex[right=1.48cm of ab] (c);
    \vertex[right=1.33cm of c] (d) {\(\overline{4}\)};
    \vertex[above=1cm of ab] (f12);
    \vertex[above right=1cm of f12] (e) {\(2\)};
    \vertex[above left=1cm of f12] (f) {\(1\)};
    \vertex[above=1.7cm of c] (g) {\(3\)};
    \diagram* {
        {[edges=fermion]
            (d)--(c)--(ab)--(a),
        },
        (ab) -- [gluon] (f12),
        (f12) -- [gluon] (f),
        (f12) -- [gluon] (e),
        (c) -- [gluon] (g),
    };
\end{feynman}
\end{tikzpicture}%
\hspace{2mm}
\begin{tikzpicture}[scale=0.75, transform shape, baseline=(a.base)]
\begin{feynman}
    \vertex (a) {\(\overline 5\)};
    \vertex[right=2cm of a] (b);
    \vertex[right=2cm of b] (c) {\(\overline{4}\)};
    \vertex[above=0.77cm of b] (d) [myblobFF] {$A$};
    \vertex[above left=1.5cm and 0.75cm of d] (g1) {\(1\)};
    \vertex[above=1.5cm of d] (g2) {\(2\)};
    \vertex[above right=1.5cm and 0.75cm of d] (g3) {\(3\)};
    \diagram* {
        (c) -- [fermion] (b) -- [fermion] (a),
        (b) -- [gluon] (d),
        (d) -- [gluon] (g1),
        (d) -- [gluon] (g2),
        (d) -- [gluon] (g3),
    };
\end{feynman}
\end{tikzpicture}
\end{equation*}
Building upon eq.~\eqref{eq:KLT}, we derive the five-point double-copy numerator as follows:
\begin{align}
    \mathcal{N}_f([123], \overline{4},\overline{5})&=\mathcal{S}[123|123]\, A_f(123, \overline 4,\overline 5)+\mathcal{S}[123|132]\, A_f(132, \overline 4,\overline 5) \, , 
\end{align}
where the elements of the KLT momentum kernel, as outlined in eq.~\eqref{Srec}, are $\mathcal{S}[123|123]=4(p_1\mdot p_2)\, (p_{12}\mdot p_3)$ and  $\mathcal{S}[123|132]=4(p_1\mdot p_2)\,( p_{1}\mdot p_3).$ 
This leads to manifestly gauge invariant and local double-copy five-point numerator
 \begin{align}
 \label{eq:fiveNum}
    \mathcal{N}_f([123],\overline{4},\overline{5})
&=\bar v\dd \Big[ {\big(p_5\mdot F_{123}\big) + {1\over 4}F_1\dd \big(p_1 \mdot F_{23}\big)\over p^2_{51}-m^2}\, \nn\\
 &-{\big(p_5\mdot F_{12}\big) + {1\over 4}F_1\dd \big(p_1 \mdot F_{2}\big)\over p^2_{51}-m^2}\dd {2(p_{12}\mdot F_{3}\mdot p_{5})+(p_{12}\mdot F_{3})\dd p_{3} \over p^2_{512}-m^2}\nn\\
 &- {\big(p_5\mdot F_{13}\big) + {1\over 4}F_1\dd \big(p_1 \mdot F_{3}\big)\over p^2_{51}-m^2}\dd {2(p_{13}\mdot F_{2}\mdot p_{5})+(p_{13}\mdot F_{2})\dd p_{2}\over p^2_{513}-m^2}\, \Big]
\dd u \, .
\end{align} 
%
Systematically obtaining such expressions for higher point amplitudes is challenging due to the intricate process of eliminating massless poles in the double-copy numerator and rewriting in terms of strings of field strength tensors. 
In the following section, we will work out an algebraic construction of numerators using Hopf algebra, which offers a direct and systematic universal approach.
\subsection{Direct algebraic construction of numerators}
We will now use Hopf algebra as an inspiration to simplify the construction of numerators and avoid the complexities of the direct computation approach. We will use pre-numerators, which follow the logic of \cite{Chen:2021chy}. 
We note that pre-numerators are not uniquely determined and still have residual freedom. However, we can eliminate it by working in a framework where we equate the left-nested numerator $\npre_{\!f}([12\ldots n\!-\!2], \sc{n{-}1},\sc{n})$ with the pre-numerator $\npre_{\!f}(12\ldots n\!-\!2, \sc{n{-}1},\sc{n})$ and fix all numerators $\npre_{\!f}(\ldots 1 \ldots, \sc{n{-}1},\sc n)$ to be zero\footnote{\footnotesize{We remark that working directly with crossing symmetric numerators will lead to the same result.}}. We next compute the pre-numerator using the fusion product
\begin{align}
    \npre_{\!f}(12\ldots n\!-\!2,\sc {n\!-\!1}, \sc n)&=\langle\!\langle T_{(1)}\star T_{(2)}\star \cdots \star T_{(n\!-\!2)}\rangle\!\rangle \,  ,
\end{align}
where we label the generators {\it {\it e.g.}} $T_{(i)}$ and $T_{(\tau_1),(\tau_2),\ldots, (\tau_r)}$ and the subscripts $\tau_{i} \subset\{1, \ldots, n\!-\!2\}$ represent an ordered partition of a subset of the external gluon indices. The evaluation map denoted by $\langle\!\langle \ldots \rangle\!\rangle$ above provides the physical gauge-invariant object that appears in the color-kinematic numerator. We can work out the fusion products by introducing quasi-shuffle operator products given by:
\begin{align}\label{eq:fusion}
    &T_{(1)}\star T_{(2)}=-T_{(12)}+T_{(1),(2)}+T_{(2),(1)}\nn \\
    & \ldots\nn \\
    &T_{(\tau_1),\ldots, (\tau_r)}\star T_{(j)}=-\sum_{i=1}^{r}T_{(\tau_1),\ldots,(\tau_i j),\ldots, (\tau_r)}+\sum_{\sigma\in \{\tau_1,\ldots,\tau_r\}\shuffle j}^r \Big(T_{(\sigma_1),\ldots, (\sigma_r),(\sigma_{r+1})}\Big) \, .
\end{align}
leading to the following explicit result for the four-point numerator:
\begin{align}
    &\npre_{\!f}(12,\sc{3},\sc 4)=\langle\!\langle T_{(1)}\star T_{(2)}\rangle\!\rangle=-\langle\!\langle T_{(12)}\rangle\!\rangle +\langle\!\langle T_{(1),(2)}\rangle\!\rangle+\langle\!\langle T_{(2),(1)}\rangle\!\rangle \, .
\end{align}
We now turn our focus to the evaluation that maps the abstract generators $T_{(\tau_1),(\tau_2),\ldots, (\tau_r)}$ to gauge invariant functions. For our massive fermion amplitudes, it reads\\[-20pt]
\begin{align}\label{eq:FPmap}
&\langle\!\langle T_{(1\tau_1),(\tau_2),\ldots,(\tau_r)}\rangle\!\rangle = \Bigg\langle\!\!\!\Bigg\langle\begin{tikzpicture}[baseline={([yshift=-0.8ex]current bounding box.center)}]\tikzstyle{every node}=[font=\small]    
   \begin{feynman}
    \vertex (a)[myblob]{};
     \vertex[left=0.8cm of a] (a0)[myblob]{};
     \vertex[right=0.8cm of a] (a2)[myblob]{};
      \vertex[right=0.8cm of a2] (a3)[myblob]{};
       \vertex[right=0.8cm of a3] (a4)[myblob]{};
       \vertex[above=0.8cm of a] (b1){$\tau_1~~~~$};
        \vertex[above=0.8cm of a2] (b2){$\tau_2$};
        \vertex[above=0.8cm of a3] (b3){$\cdots$};
         \vertex[above=0.8cm of a4] (b4){$\tau_r$};
         \vertex[above=0.8cm of a0] (b0){$1~~$};
       \vertex [above=0.8cm of a0](j1){$ $};
    \vertex [right=0.2cm of j1](j2){$ $};
    \vertex [right=0.6cm of j2](j3){$ $};
    \vertex [right=0.4cm of j3](j4){$ $};
    \vertex [right=0.8cm of j4](j5){$ $};
      \vertex [right=0.2cm of j5](j6){$ $};
    \vertex [right=0.6cm of j6](j7){$ $};
     \vertex [right=0.0cm of j7](j8){$ $};
    \vertex [right=0.8cm of j8](j9){$ $};
   	 \diagram*{(a)--[very thick](a0),(a)--[very thick](a2),(a2)--[very thick](a3), (a3)--[very thick](a4),(a0) -- [thick] (j1),(a) -- [thick] (j2),(a)--[thick](j3),(a2) -- [thick] (j4),(a2)--[thick](j5),(a4) -- [thick] (j8),(a4)--[thick](j9)};
    \end{feynman}  
  \end{tikzpicture}\Bigg\rangle\!\!\!\Bigg\rangle\\
 &= \bar v\dd {H_{1\tau_1}\over p^2_{n1}-m^2}\dd {p_{n1\tau_1}{-}m\over p^2_{n1\tau_1}-m^2}\, \dd (p_{\Theta_L(\tau_{2})}\mdot F_{\tau_2})\dd\cdots\dd {p_{n1\tau_1\ldots \tau_{r-1}}{-}m \over p^2_{n1\tau_1\cdots \tau_{r-1}}-m^2 }\dd (p_{\Theta_L(\tau_{r})}\mdot F_{\tau_r})\dd u \,, \nn
\end{align}\\[-15pt]
where we have $\la T_{(1),\ldots}\ra=\la T_{\ldots,(1\tau_i),\ldots}\ra =0$ and 
\begin{align}\label{eq:G1}
	H_{1\tau}\equiv p_n\mdot F_{1\tau} + {1\over 4}F_1\dd \big(p_1 \mdot F_{\tau}\big).
\end{align}
The symbol $\Theta_L(\tau_{r})$ denotes all indices in $(1\tau_1)\cup (\tau_2)\cup\cdots \cup (\tau_{r-1})$ that are smaller than the first element in $\tau_{r}$. We can write a closed-form color-kinematic numerator as follows: \\[-15pt]
\begin{align}\label{eq:closedformFP}
    \npre_{\!f}(12\ldots n\!-\!2, \sc{n\!-\!1},\sc n)=\sum_{\rm ordered\  partitions}({-}1)^{n+r} \langle\!\langle T_{(1\tau_1),(\tau_2),\ldots,(\tau_r)}\rangle\!\rangle \,,
\end{align}\\[-15pt]
where sum is over ordered partitions of the set $\{2,3,...,n\!-\!2\}$ into $r$ nonempty subsets\footnote{More examples of color-kinematic numerators and a {\tt Mathematica} package can be found at  {\href{https://github.com/AmplitudeGravity/kinematicHopfAlgebra}{{\it \blue KiHA5.0} GitHub repository}}~\cite{ChenGitHub}.}. We can now expand this sum as follows.\\[-15pt] %
\begin{align}\label{eq:rec}
   &\npre_{\!f}(1\ldots n\!-\!2,\sc {n\!-\!1}, \sc n)=(-1)^{n\!-\!3} \ \bar v\dd { H_{1\ldots n\!-\!2}\over p^2_{n1}-m^2}\dd u\nn \\
   &+\sum_{\tau_R\subset \{2,\ldots,n\!-\!2\}} (-1)^{|\tau_R|} \ \bar v \dd \Big[ {{\cal X}\big(1\rho{(\tau_R)}\big)\dd (p_{n1\rho{(\tau_R)}}{-}m)\dd (p_{\Theta_L(\tau_R)}\mdot F_{\tau_R}) \over p^2_{n1\rho{(\tau_R)}}-m^2}\Big]\dd u \,,
\end{align}
where we have defined $|\tau_R|$ as the number of elements in the set $\tau_R$, and $\rho{(\tau_R)}$ as the complement legs in $\{2,\ldots, n\!-\!2\}$. ${\cal X}(1\rho{(\tau_R)})$ represents a particular Dirac matrix product of strengthen tensors and massive propagators for the set of gluon legs $\{1\rho{(\tau_R)}\}$ that correspond to a color-kinematic numerator that is stripped of incoming and outgoing external fermion states. 
From this, we introduce a factorization on an internal massive leg denoted by $I$ and give the expression,\\[-15pt]
\begin{align}\label{eq:rec2}
  \npre_{\!f}(1\ldots n\!-\!2,&\sc {n\!-\!1}, \sc n) =(-1)^{n\!-\!3} \ \bar v\dd { H_{1\ldots n\!-\!2}\over p^2_{n1}-m^2}\dd u  \nn\\
   &+\sum_{\tau_R\subset \{2,\ldots,n\!-\!2\}} \sum_{S_{I}}(-1)^{|\tau_R|} \ \npre_{\!f}(1\rho{(\tau_R)},\sc I,\sc n)\,\bar v_I\dd {  (p_{\Theta_L(\tau_R)}\mdot F_{\tau_R}) \over p^2_{n1\rho{(\tau_R)}}-m^2} \dd u \, . \nn
\end{align}
where $S_{I}$ denotes the possible spin states of $I$ and where we have used in the last step that $\bar v\dd {\cal X}(1\rho{(\tau_R)})\dd u_I \equiv \npre_{\!f}(1\rho{(\tau_R)},\sc I,\sc n)$. 

To prove the recursive form shown in eq.~\eqref{eq:rec}, we start by considering the factorization behavior on the massive propagator that involves only one massless particle $j$ on the right-hand side. We then take the collinear limit, in legs $j$ and $n\!-\!1$,
\begin{align}
&\npre_{\!f}(1\ldots n\!-\!2, \sc {n{-}1},\sc n)
\longrightarrow
\begin{tikzpicture}[baseline={([yshift=-0.8ex]current bounding box.center)}]
\tikzstyle{every node}=[font=\small]  
\tikzstyle{mybox} = [rectangle, draw=black, fill=red, minimum size=0.3cm]
\tikzstyle{mydot} = [circle, draw=black, fill=black, minimum size=3pt,inner sep=0pt]
\begin{feynman}
 \vertex (a)[mybox]{};
 \vertex[left=0.8cm of a] (a0){};
 \vertex[right=0.8cm of a] (cut){};
    \vertex[below=0.2cm of cut] (I){$I$};
 \vertex[below=0.2cm of a0] (n){$n$};
 \vertex[right=2.4cm of a] (a1){};
 \vertex[right=1.6cm of a] (j0)[mydot]{};
 \draw[red, thick] ($(cut) + (-0.1,-0.1)$) -- ($(cut) + (0.1,0.1)$);
 \draw[red, thick] ($(cut) + (-0.1,0.1)$) -- ($(cut) + (0.1,-0.1)$colin);
 \vertex[above=0.9cm of j0] (j1){$j$};
 \vertex[below=0.2cm of a1] (n1){$n{-}1$};
 \vertex[above=0.9cm of a] (b){$\,\,1\,\,\,\,\,\rho{(j)}$};
 \vertex[above=0.8cm of a0] (b0){$ $};
 \vertex[right=0.4cm of b0] (b11){$ $};
 \vertex[right=0.8cm of b11] (b12){$ $};
 \diagram*{(a)--[very thick](a0),(a)--[very thick](a1),(a) -- [thick] (b11),(a)--[thick](b12),(j1)--[thick](j0)};
\end{feynman}  
\end{tikzpicture}\nn\\
    &= \sum_{S_{I}}{\sum_{\beta_L\in S_{n{-}4}}}\,\mathcal{S}[1\rho{(j)}|1\beta_L]\,A(1\beta_L,\sc I,\sc n)\,(p_{\Theta_L(j)}\mdot p_j) \, A(j,\sc{n{-}1},\sc I)\nn\\ &=\sum_{S_{I}}\,\npre_f(1\rho{(j)},\sc I,\sc n)\, (p_{\Theta_L(j)}\mdot p_j) \, \npre_f(j,\sc{n{-}1},\sc I) \, . 
\end{align}
We have in the above equation utilized eq.~\eqref{eq:KLT} and  eq.~\eqref{Srec}.
Using the completeness relation for Dirac spinors and explicitly writing the expression for the fermionic propagator, the sum over intermediate states yields the following after evaluating the three-point numerator on the right side of the cut:
\begin{align}\label{factorisation last leg}
    &\npre_{\!f}(1\ldots n\!-\!2, \sc {n\!-\!1},\sc n)\rightarrow (p_{\Theta_L(j)}\mdot p_j)\, \bar v \dd {\cal X}(1\rho{(j)})\dd ({p}_{n1\rho{(j)}}{-}m) \dd \veps_j \dd u \, .
\end{align}
We can now deduce a manifest gauge invariant part of the numerator that replicates this factorization behavior. 
\begin{equation}
    \begin{aligned}\label{RECN1}
    \npre^{(1)}(1\rho{(j)}|j, \overline {n\!-\!1},\overline n)
    &= -{ \bar v \dd {\cal X}(1\rho{(j)})\dd (p_{n1\rho{(j)}}-m) \dd (p_{\Theta_L(j)}\mdot F_j)\dd u \over p_{n1\rho{(j)}}^2-m^2}\, ,\\
\end{aligned}
\end{equation}
where $\rho{(j)}|j$ indicates that the particle to the right of $\rho{(j)}$ is $j$.  
Subtracting all the possible one-leg contributions from the complete color-kinematic numerator, we obtain a remainder quantity,
\begin{align}\label{eq:diff1}
\npre_f(1\ldots n\!-\!2, \sc {n\!-\!1},\sc n)-\sum_{j} \npre^{(1)}(1\rho{(j)}|j, \sc {n\!-\!1},\sc n) \, .
\end{align}
By construction, it should also manifestly gauge invariant, free of any massive poles, and contain $n\!-\!3$ massless legs on the left-hand side (since the pieces we have subtracted contain the complete factorization behavior for those configurations). Next, we consider a massive cut containing two massless legs on the right-hand side
\begin{align}
\begin{tikzpicture}[baseline={([yshift=-0.8ex]current bounding box.center)}]
\tikzstyle{every node}=[font=\small]  
\tikzstyle{mybox} = [rectangle, draw=black, fill=red, minimum size=0.3cm]
\tikzstyle{mydot} = [circle, draw=black, fill=black, minimum size=3pt,inner sep=0pt]
\begin{feynman}
 \vertex (a)[mybox]{};
 \vertex[left=0.8cm of a] (a0){};
 \vertex[right=0.8cm of a] (cut){};
    \vertex[below=0.2cm of cut] (I){$I$};
 \vertex[below=0.2cm of a0] (n){$n$};
 \vertex[right=2.4cm of a] (a1){};
 \vertex[right=1.6cm of a] (j0)[mydot]{};
 \draw[red, thick] ($(cut) + (-0.1,-0.1)$) -- ($(cut) + (0.1,0.1)$);
 \draw[red, thick] ($(cut) + (-0.1,0.1)$) -- ($(cut) + (0.1,-0.1)$colin);
 \vertex[above=0.9cm of j0] (j1){};
  \vertex[left=0.2cm of j1] (j11){$j_1$};
 \vertex[right=0.2cm of j1] (j12){$j_2$};
 \vertex[below=0.2cm of a1] (n1){$n{-}1$};
 \vertex[above=0.9cm of a] (b){$\,\,1\,\,\,\,\,\rho{(j_1j_2)}$};
 \vertex[above=0.8cm of a0] (b0){$ $};
 \vertex[right=0.4cm of b0] (b11){$ $};
 \vertex[right=0.8cm of b11] (b12){$ $};
 \diagram*{(a)--[very thick](a0),(a)--[very thick](a1),(a) -- [thick] (b11),(a)--[thick](b12),(j11)--[thick](j0),(j12)--[thick](j0)};
\end{feynman}  
\end{tikzpicture} \, .
\end{align}
 Then the remainder  eq.~\eqref{eq:diff1} in this case tends to 
\begin{align}\label{factorisation two legs}
    (p_{\Theta_L(j_1)}\mdot p_{j_1}) \bar v \dd {\cal X}(1\rho{(j_1j_2)})\dd (p_{n1\rho{(j_1j_2)}}{-}m) \dd X(j_1 j_2) \dd u \, .
\end{align}
where $\rho(j_1j_2)$ is the complement of $j_1,j_2$ in the set $1,2,\ldots, n-2$. The symbol $X(j_1 j_2)$, whose exact form is unimportant at this point, is a function of the polarizations and momenta of the external gluon legs associated with labelings $j_1,j_2$. 

Now, subtracting the pieces that contribute to further poles in $X(j_1j_2)$ yields a rank one tensor structure to the right of the propagator $(p_{n1 \rho{(j_1j_2)} }{-}m)$. Indeed, any higher rank tensor had to be generated by a Feynman diagram containing additional fermionic propagators, and the only way of canceling the pole would be by using the Dirac matrix algebra, which in turn lowers the tensor rank. To determine the gauge invariant tensor structure, one can fit the factorization behavior in the heavy-mass regime by parameterizing $p_n=m\, v$, $p_{n\!-\!1} = -m\, v - q$ and taking $m\to\infty$. Then the corresponding remainder quantity yields \cite{Brandhuber:2021bsf}
\begin{align}\label{eq:heftDiff}
    \npre(1\rho{(j_1j_2)},v)\times p_{\Theta_L(j_1)}\mdot F_{j_1j_2}\mdot v \, .
\end{align}
Comparing the heavy mass limit of  eq.~\eqref{eq:diff1} with  eq.~\eqref{eq:heftDiff}, the piece of the numerator that contributes exclusively to the two-particle cut is found to be:
\begin{align}
     &\npre^{(2)}(1\rho{(j_1j_2)} |j_1 j_2, \sc {n\!-\!1},\sc n)= { \bar v \dd {\cal X}(1\rho{(j_1j_2)})\dd (p_{n1\rho{(j_1j_2)}}{-}m)\dd (p_{\Theta_L(j_1)}\mdot F_{j_1j_2})\dd u \over p^2_{n1\rho{(j_1j_2)}}-m^2} \, .
\end{align}
Subtracting these new contributions to the color-kinematic numerator results in an object that is free of one- and two-leg poles:
\begin{align}
    &\npre_f(1\ldots n\!-\!2, \sc {n\!-\!1},\sc n)-\sum_{j} \npre^{(1)}(1\rho{(j)}|j, \sc {n\!-\!1},\sc n)-\sum_{j_1\,<\,j_2} \npre^{(2)}(1\rho{(j_1j_2)}|j_1 j_2, \sc {n\!-\!1},\sc n) \, .
\end{align}

The procedure can be repeated recursively for cuts containing any number $i<n\!-\!3$ of legs on the right-hand side:
\begin{align}
     &\npre^{(i)}(1\rho{(\tau_R)}|\tau_R, \sc {n\!-\!1},\sc n)=(-1)^{i} { \bar v \dd {\cal X}(1\rho{(\tau_R)})\dd (p_{n1\rho{(\tau_R)}}{-}m) \dd (p_{\Theta_L(\tau_R)}\mdot F_{\tau_R})\dd u \over p^2_{n1\rho{(\tau_R)}}-m^2} \, .
\end{align}
For the last cut with only leg one on the left, the expression contains an additional term, 
\begin{align}
    \npre^{(n\!-\!3)}(1|2\ldots n\!-\!2, \sc {n\!-\!1},\sc n)
    &=(-1)^{n\!-\!3} \ \bar v\dd {(p_n\mdot F_{12\ldots n\!-\!2}) + \frac{1}{4}{F_1} \dd (p_1 \mdot F_{2\ldots n\!-\!2})\over p^2_{n1}-m^2}\dd u \nn\\
    &=(-1)^{n\!-\!3} \ \bar v\dd {H_{12\ldots n\!-\!2}\over p^2_{n1}-m^2}\dd u \, .
\end{align}

By repeating our previous argument, we see that subtracting all these contributions, 
\begin{align}
    \npre_f(1\ldots n\!-\!2, \sc {n\!-\!1},\sc n)-\npre^{(1)}-\npre^{(2)}-\cdots- \npre^{(n\!-\!3)}\,,
\end{align}
leads to an object that contains no poles, {\it {\it i.e.},} a polynomial piece. Through the use of dimensional analysis, it becomes apparent that a color-kinematic numerator must include the momenta $p_i$ raised to an overall power of $2$ and the polarization vector $\varepsilon_i$ raised to an overall power of $3$. It has been proven that no polynomial expression, which is manifestly gauge invariant, can satisfy these power requirements. This implies that the remainder must be equal to zero. Therefore, the recursive form presented in eq.~\eqref{eq:rec} is valid, and as a result, the evaluation map in eq.~\eqref{eq:FPmap} is correct.

\section{An alternative evaluation map for generating simpler color-kinematic numerators}\label{sec:map2}
The previous section's evaluation map did not use the massive on-shell state condition. This section explores using this auxiliary constraint to provide an alternative graphical evaluation map. This map removes expression redundancies and further simplifies the intricate Dirac matrix product in the numerators. We will start by considering numerator examples at four, five, and six points, which will form a basis for universal and more straightforward graphical rules for evaluating numerator expressions. 

\subsection{Four-point color-kinematic numerator}
Our first example is the four-point color-kinematic numerator. Here, we get directly from the algebraic construction 
 \begin{align}
    {\mathcal{N}}_{\!f}(12,\overline{3},\overline{4})=& \la T_{(12)}\ra =\bar v\dd  {H_{12}\over p^2_{41}-m^2}\, 
\dd u \, ,
\end{align} 
and an expression, which uses the on-shell condition and $H_{12}=p_4\mdot F_{12} + {1\over 4}F_1\dd \big(p_1 \mdot F_{2}\big)$ from the general definition eq.~\eqref{eq:G1}. 

\subsection{Five-point color-kinematic numerator}
At five-point, the algebraic construction yields:
   \begin{align}
    {\mathcal{N}}_{\!f}(123,\overline{4},\overline{5})=& \la T_{(123)}-T_{(12),(3)}-T_{(13),(2)}\ra \nn \\
    =&\bar v\dd \Big[ {H_{123}\over p^2_{51}-m^2}\, 
 -{H_{12}\over p^2_{51}-m^2}\dd {p_{512}-m\over p^2_{512}-m^2}\dd (p_{12}\mdot F_3)\,\nn\\
 &~~~~~~~~~~~~~~~~~~-{H_{13}\over p^2_{51}-m^2}\dd {p_{513}-m\over p^2_{513}-m^2}\dd (p_{1}\mdot F_2)\, \Big]
\dd u \, \nn\\
&=\bar v\dd \Big[ {H_{123}\over p^2_{51}-m^2}\, -{H_{12}\over p^2_{51}-m^2}\dd {2(p_{12}\mdot F_{3}\mdot p_{5})+(p_{12}\mdot F_{3})\dd p_{3} \over p^2_{512}-m^2}\nn\\
 &~~~~~~~~~~~~~~~~~~~~~~- {H_{13}\over p^2_{51}-m^2}\dd {2(p_{13}\mdot F_{2}\mdot p_{5})+(p_{13}\mdot F_{2})\dd p_{2}\over p^2_{513}-m^2}\, \Big]
\dd u \, .
\end{align} 
In the last step, we use the Clifford algebra and the on-shell condition to reduce the rank of the two tensors to a scalar product and obtain a simpler tensor rank. The {\it H} functions are  
\begin{align}
H_{12}&=p_5\mdot F_{12} + {1\over 4}F_1\dd \big(p_1 \mdot F_{2}\big)\, ,\nn\\
	H_{13}&=p_5\mdot F_{13} + {1\over 4}F_1\dd \big(p_1 \mdot F_{3}\big)\, , \nn\\
	H_{123}&=p_5\mdot F_{123} + {1\over 4}F_1\dd \big(p_1 \mdot F_{23}\big)\, .
\end{align}
\subsection{Six-point color-kinematic numerator}
Finally, we consider the six-point color-kinematic numerator. Our starting point is the algebraic construction of the numerator, which reads,
\begin{align}\label{N6}
    \npre_{\!f}(1234,\sc 5,\sc 6)&=\la 
    -T_{\text{(12)},\text{(3)},\text{(4)}}-T_{\text{(12)},\text{(4)},\text{(3)}}+T_{\text{(12)},\text{(34)}}
    -T_{\text{(13)},\text{(2)},\text{(4)}}\nn\\
    &
    -T_{\text{(13)},\text{(4)},\text{(2)}}+T_{\text{(13)},\text{(24)}}-T_{\text{(14)},\text{(2)},\text{(3)}}
    -T_{\text{(14)},\text{(3)},\text{(2)}}\nn\\
    &+T_{\text{(14)},\text{(23)}}
    +T_{\text{(123)},\text{(4)}}+T_{\text{(124)},\text{(3)}}
    +T_{\text{(134)},\text{(2)}}-T_{\text{(1234)}}\ra \, .
\end{align}
$T_{(1234)}$ is unchanged. $T_{\text{(123)},\text{(4)}},T_{\text{(124)},\text{(3)}},T_{\text{(134)},\text{(2)}}$ are simplified as in the five point case. There are new non-trivial cancelations for the terms associated with the evaluations of
$\la -T_{\text{(12)},\text{(3)},\text{(4)}}-T_{\text{(12)},\text{(4)},\text{(3)}}+T_{\text{(12)},\text{(34)}}\ra$, $\la -T_{\text{(13)},\text{(2)},\text{(4)}}-T_{\text{(13)},\text{(4)},\text{(2)}}+T_{\text{(13)},\text{(24)}}\ra$ and $\la -T_{\text{(14)},\text{(2)},\text{(3)}}-T_{\text{(14)},\text{(3)},\text{(2)}}+T_{\text{(14)},\text{(23)}}\ra$. 
Writing out the first one of these terms, we get:
\begin{align}
&\la -T_{\text{(12)},\text{(3)},\text{(4)}}-T_{\text{(12)},\text{(4)},\text{(3)}}+T_{\text{(12)},\text{(34)}}\ra \nn\\
&= -\Bigg\langle\!\!\!\Bigg\langle\begin{tikzpicture}[baseline={([yshift=-0.8ex]current bounding box.center)}]\tikzstyle{every node}=[font=\small]    
   \begin{feynman}
    \vertex (a)[myblob]{};
     \vertex[left=0.8cm of a] (a0)[myblob]{};
     \vertex[right=0.8cm of a] (a2)[myblob]{};
       \vertex[right=0.8cm of a2] (a4)[myblob]{};
       \vertex[above=0.8cm of a] (b1){$2~~~~~~~~$};
        \vertex[above=0.8cm of a2] (b2){$3$};
         \vertex[above=0.8cm of a4] (b4){$4$};
         \vertex[above=0.8cm of a0] (b0){$1~~$};
       \vertex [above=0.8cm of a0](j1){$ $};
    \vertex [right=0.2cm of j1](j2){$ $};
    \vertex [right=1.4cm of j2](j4){$ $};
      \vertex [right=0.8cm of j4](j6){$ $};
     \vertex [right=0.0cm of j6](j8){$ $};
   	 \diagram*{(a)--[very thick](a0),(a)--[very thick](a2),(a2)--[very thick](a3), (a3)--[very thick](a4),(a0) -- [thick] (j1),(a) -- [thick] (j2),(a2) -- [thick] (j4),(a4) -- [thick] (j8)};
    \end{feynman}  
  \end{tikzpicture}\Bigg\rangle\!\!\!\Bigg\rangle-\Bigg\langle\!\!\!\Bigg\langle\begin{tikzpicture}[baseline={([yshift=-0.8ex]current bounding box.center)}]\tikzstyle{every node}=[font=\small]    
   \begin{feynman}
    \vertex (a)[myblob]{};
     \vertex[left=0.8cm of a] (a0)[myblob]{};
     \vertex[right=0.8cm of a] (a2)[myblob]{};
       \vertex[right=0.8cm of a2] (a4)[myblob]{};
       \vertex[above=0.8cm of a] (b1){$2~~~~~~~~$};
        \vertex[above=0.8cm of a2] (b2){$4$};
         \vertex[above=0.8cm of a4] (b4){$3$};
         \vertex[above=0.8cm of a0] (b0){$1~~$};
       \vertex [above=0.8cm of a0](j1){$ $};
    \vertex [right=0.2cm of j1](j2){$ $};
    \vertex [right=1.4cm of j2](j4){$ $};
      \vertex [right=0.8cm of j4](j6){$ $};
     \vertex [right=0.0cm of j6](j8){$ $};
   	 \diagram*{(a)--[very thick](a0),(a)--[very thick](a2),(a2)--[very thick](a3), (a3)--[very thick](a4),(a0) -- [thick] (j1),(a) -- [thick] (j2),(a2) -- [thick] (j4),(a4) -- [thick] (j8)};
    \end{feynman}  
  \end{tikzpicture}\Bigg\rangle\!\!\!\Bigg\rangle+\Bigg\langle\!\!\!\Bigg\langle\begin{tikzpicture}[baseline={([yshift=-0.8ex]current bounding box.center)}]\tikzstyle{every node}=[font=\small]    
   \begin{feynman}
    \vertex (a)[myblob]{};
     \vertex[left=0.8cm of a] (a0)[myblob]{};
     \vertex[right=0.8cm of a] (a2)[myblob]{};
       \vertex[above=0.8cm of a] (b1){$2~~~~~~~~$};
        \vertex[above=0.8cm of a2] (b2){$34$};
         \vertex[above=0.8cm of a0] (b0){$1~~$};
       \vertex [above=0.8cm of a0](j1){$ $};
    \vertex [right=0.2cm of j1](j2){$ $};
    \vertex [right=1.2cm of j2](j4){$ $};
    \vertex [right=0.4cm of j4](j5){$ $};
   	 \diagram*{(a)--[very thick](a0),(a)--[very thick](a2), (a0) -- [thick] (j1),(a) -- [thick] (j2),(a2) -- [thick] (j4),(a2) -- [thick] (j5)};
    \end{feynman}  
  \end{tikzpicture}\Bigg\rangle\!\!\!\Bigg\rangle\nn\\
&=-\frac{\bar{v}\dd H_{12} \dd (p_{612}-m)\dd \left(p_{12}\mdot F_3\right)\dd (p_{6123}-m)\dd \left(p_{123}\mdot F_4\right)\dd u}{(p_{61}^2-m^2) (p_{612}^2-m^2) (p_{6123}^2-m^2)}\nn\\
&-\frac{\bar{v}\dd H_{12}\dd (p_{612}-m)\dd \left(p_{12}\mdot F_4\right)\dd (p_{6124}-m)\dd \left(p_{12}\mdot F_3\right)\dd u}{(p_{61}^2-m^2) (p_{612}^2-m^2) (p_{6124}^2-m^2)}\nn\\
&
+\frac{\bar{v}\dd H_{12}\dd (p_{612}-m)\dd \left(p_{12}\mdot F_3\mdot F_4\right)\dd u}{(p_{61}^2-m^2) (p_{612}^2-m^2)} \, ,
\end{align}
where 
\begin{align}
H_{12}&=p_6\mdot F_{12} + {1\over 4}F_1\dd \big(p_1 \mdot F_{2}\big)\,  .
\end{align}
Terms involving two propagators yield
\begin{align}\label{6pteg}
   &\bar{v}\dd\frac{ H_{12}}{p_{61}^2-m^2}\dd\left[\frac{\left(p_{12}\mdot F_3\right)\dd \left(p_{123}\mdot F_4\right)}{(p_{612}^2-m^2) }+\frac{ \left(p_{12}\mdot F_4\right)\dd  \left(p_{12}\mdot F_3\right)}{(p_{612}^2-m^2)}+\frac{ (p_{612}-m)\dd \left(p_{12}\mdot F_{34}\right)}{(p_{612}^2-m^2) }\right]\dd u\nn\\
   =&\bar{v}\dd\frac{ H_{12}}{p_{61}^2-m^2}\dd\Big[\frac{ \left(p_{12}\mdot F_3\right)\dd \left(p_{123}\mdot F_4\right)}{(p_{612}^2-m^2)  }+\frac{ \left(p_{12}\mdot F_4\right)\dd  \left(p_{12}\mdot F_3\right)}{(p_{612}^2-m^2) }
   +\frac{ \left(p_{12}\mdot F_{34}\right)\dd p_{34}}{(p_{612}^2-m^2)  }\nn\\
 &~~~~~~~~~~~~~~~~~+\frac{2\left(p_{12}\mdot F_{34} \mdot p_{612}\right)}{(p_{612}^2-m^2) }\Big]\dd u  \, .
\end{align}
The terms contained within the square brackets on the right-hand side we rephrase further as follows:
\begin{align}
    &\frac{2\left(p_{12}\mdot F_3\mdot F_4 \mdot p_{612}\right)}{(p_{612}^2-m^2) }+
    \frac{ \left(p_{12}\mdot F_3\right)\dd \left(p_{3}\mdot F_4\right)}{(p_{612}^2-m^2) }
    +\frac{ \left(p_{12}\mdot F_3\mdot F_4\right)\dd p_{34}}{(p_{612}^2-m^2) }
   \nn\\
   & +\frac{ \left(p_{12}\mdot F_3\right)\dd \big(p_{12}\mdot F_4\big)}{(p_{612}^2-m^2) }
    +\frac{ \left(p_{12}\mdot F_4\right)\dd  \left(p_{12}\mdot F_3\right)}{(p_{612}^2-m^2) } \, ,
\end{align}
and reexpress by employing the identity 
\begin{align}
  \frac{ \left(p_{12}\mdot F_3\right)\dd \big(p_{12}\mdot F_4\big)}{(p_{612}^2-m^2) }
    +\frac{ \left(p_{12}\mdot F_4\right)\dd  \left(p_{12}\mdot F_3\right)}{(p_{612}^2-m^2) }=
    -\frac{2\left(p_{12}\mdot F_3\mdot F_4 \mdot p_{12}\right)}{(p_{612}^2-m^2) } \, .
\end{align}
It leads to the following simplified expression:
\begin{align}
    &\bar{v}\dd\frac{ H_{12}}{p_{61}^2-m^2}\dd\left[\frac{2\left(p_{12}\mdot F_3\mdot F_4 \mdot p_{6}\right)}{(p_{612}^2-m^2) }
    +\frac{ \left(p_{12}\mdot F_3\right)\dd  \left(p_{3}\mdot F_4\right)}{(p_{612}^2-m^2)}
    +\frac{ \left(p_{12}\mdot F_3\mdot F_4\right)\dd p_{34}}{(p_{612}^2-m^2) }
    \right]\dd u \, .
\end{align}
From this expression, we make the following observations. It is manifest that, in the simplified form, all the mass-dependent contributions are gone in the numerator, and the massive momentum-dependent terms only appear in the scalar part. Fully accounting for the on-shell conditions allows us a much-simplified expression.

  It follows from above examples, that the function $H_{1\tau}$ in the evaluation map of the first component of the algebraic generator is unchanged. Other components are reduced to  
\begin{align}
  H_{j\tau}\big(p_\rho,p_\omega\big)&=2 p_{\rho}\mdot F_{j\tau}\mdot p_{\omega}+
\sum_{\sigma_1\in \tau}\limits \Big(p_{\rho}\mdot  F_{j\sigma_1}\Big)\dd\Big( p_{\Theta'_L(\sigma_2)}\mdot F_{\sigma_2}\Big) \, , \, \text{if}~ j>1\, .
\end{align}
The expression involves the summation of all possible orderings of $\tau$, denoted as $\sigma_1\sigma_2\equiv \tau$, here, $\Theta'_L(\sigma_2)$ represents all the indices in $j\sigma_1$ that come before the index of $\sigma_2$ in the canonical ordering. When $\sigma_2$ is an empty set, $\Theta'_L$ includes every index in $j\tau$. In this case, $F_{\sigma_2}$ reduces to an identity matrix. For example 
\begin{align}
	H_{j}\big(p_\rho,p_\omega\big)&=2 p_{\rho}\mdot F_{j\tau}\mdot p_{\omega}+ \Big(p_{\rho}\mdot  F_{j}\Big)\dd p_{j}\,,\nn\\
	H_{jk}\big(p_\rho,p_\omega\big)&=2 p_{\rho}\mdot F_{jk}\mdot p_{\omega}+ \Big(p_{\rho}\mdot  F_{jk}\Big)\dd p_{jk}+ \Big(p_{\rho}\mdot  F_{j}\Big)\dd \Big(p_{j}\mdot  F_{k} \Big)\, .
\end{align}

\subsection{General color-kinematic numerator}
Armed with the understanding from the above examples, we will now introduce a more general and universal formalism. 
We have the following simplified evaluation map for numerators:
\begin{align}\label{eq:J}
&\langle\!\langle T_{(1\tau_1),(\tau_2),\ldots, (\tau_r)}\rangle\!\rangle \equiv \Bigg\langle\!\!\!\Bigg\langle\begin{tikzpicture}[baseline={([yshift=-0.8ex]current bounding box.center)}]\tikzstyle{every node}=[font=\small]    
   \begin{feynman}
    \vertex (a)[myblob]{};
     \vertex[left=0.8cm of a] (a0)[myblob]{};
     \vertex[right=0.8cm of a] (a2)[myblob]{};
      \vertex[right=0.8cm of a2] (a3)[myblob]{};
       \vertex[right=0.8cm of a3] (a4)[myblob]{};
       \vertex[above=0.8cm of a] (b1){$\tau_1~~~~$};
        \vertex[above=0.8cm of a2] (b2){$\tau_2$};
        \vertex[above=0.8cm of a3] (b3){$\cdots$};
         \vertex[above=0.8cm of a4] (b4){$\tau_r$};
         \vertex[above=0.8cm of a0] (b0){$1~~$};
       \vertex [above=0.8cm of a0](j1){$ $};
    \vertex [right=0.2cm of j1](j2){$ $};
    \vertex [right=0.6cm of j2](j3){$ $};
    \vertex [right=0.4cm of j3](j4){$ $};
    \vertex [right=0.8cm of j4](j5){$ $};
      \vertex [right=0.2cm of j5](j6){$ $};
    \vertex [right=0.6cm of j6](j7){$ $};
     \vertex [right=0.0cm of j7](j8){$ $};
    \vertex [right=0.8cm of j8](j9){$ $};
   	 \diagram*{(a)--[very thick](a0),(a)--[very thick](a2),(a2)--[very thick](a3), (a3)--[very thick](a4),(a0) -- [thick] (j1),(a) -- [thick] (j2),(a)--[thick](j3),(a2) -- [thick] (j4),(a2)--[thick](j5),(a4) -- [thick] (j8),(a4)--[thick](j9)};
    \end{feynman}  
  \end{tikzpicture}\Bigg\rangle\!\!\!\Bigg\rangle\nn\\
  &=\bar v\dd {H_{1\tau_1}\over p^2_{n1}-m^2}\dd {H_{\tau_2}\big(p_{\Theta_L(\tau_{2})},p_{\Theta_R(\tau_{2})}\big)\over p^2_{n1\tau_1}-m^2}\, \dd\cdots\dd {H_{\tau_r}\big(p_{\Theta_L(\tau_{r})},p_{\Theta_R(\tau_{r})}\big)\over p^2_{n1\tau_1\ldots \tau_{r-1}}-m^2}\dd u \,.
\end{align}
We obtain this map by analyzing the color-kinematic numerator pattern observed in the above examples provided by eq.~\eqref{eq:FPmap} using the additional on-shell conditions. 
We will also use the convenient musical diagrams \cite{Brandhuber:2021bsf} to ease the computation of numerators. The musical diagrams  take the form
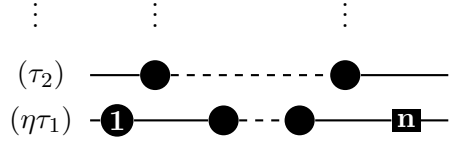
\begin{equation*}\label{eq:musicNonzero}
\begin{tikzpicture}[baseline={([yshift=-0.8ex]current bounding box.center)}]\tikzstyle{every node}=[font=\small]    
   \begin{feynman}
    \vertex (l1)[]{}; 
    \vertex [below=0.6cm of l1](ls)[]{}; 
    \vertex [above=0.9cm of l1](lg)[]{$\vdots~~~~~~~~~$}; 
     \vertex [right=1.0cm of lg](lg3)[]{$\vdots$}; 
    \vertex [right=3.5cm of lg](lg2)[]{$\vdots$}; 
    \vertex [left=0.5cm of ls](rms)[]{$(\eta\tau_1)$};
    \vertex [right=5.cm of ls](rs){}; 
    \vertex [right=4.3cm of ls](vphi)[myblob2]{\white\textbf\small $\mathbf{n}$}; 
    \vertex [left=0.5cm of l1](rm1)[]{$(\tau_2)$};
    \vertex [right=5.cm of l1](r1)[]{}; 
    \vertex [right=0.5cm of ls](v1)[sb]{\white\textbf\small $\mathbf{1}$};
     \vertex [right=1.9cm of ls](v11)[sb]{\white\textbf\small $\mathbf{~\,~}$};
     \vertex [right=2.9cm of ls](v12)[sb]{\white\textbf\small $\mathbf{~\,~}$};
    \vertex [right=1.cm of l1](v3)[sb]{\white\textbf\small $\mathbf{~\,~}$};
    \vertex [right=3.5cm of l1](v4)[sb]{\white\textbf\small $\mathbf{~\,~}$};
   	 \diagram*{(l1)- -[thick](v3)- -[scalar,thick](v4)--[thick](r1),(ls)--[thick](v1)--[thick](v11)--[scalar,thick](v12)--[thick](vphi)--[thick](rs)};
    \end{feynman}  
  \end{tikzpicture} 
\end{equation*}
In a given diagram, the reference leg indices are located at the bottom line, and we sum all possible positions of the non-reference legs. For a given diagram, each horizontal line contains an argument of an {\it H} function required for the corresponding numerator. (We use the notation $\Theta_{L/R}(\tau_i)$ that refers to the set of all the left-lower/right-lower indices of $\tau_i$. 
This definition of $\Theta_{L}(\tau_i)$ is consistent with the $\Theta_{L}$ function in the definition of gravity KLT momentum kernel used in eq.~\eqref{Srec}). 

Below, we illustrate using musical diagrams to evaluate abstract generator terms. As an example, we can consider $\langle\!\langle T_{(16),(235),(4)} \rangle\!\rangle$, which appears in the eight-point color-kinematic numerator.
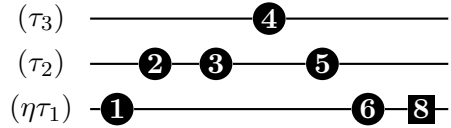
\begin{equation*}
\begin{tikzpicture}[baseline={([yshift=-0.8ex]current bounding box.center)}]\tikzstyle{every node}=[font=\small]    
   \begin{feynman}
    \vertex (l1)[]{}; 
    \vertex [below=0.4cm of l1](ls)[]{}; 
    \vertex [above=0.2cm of l1](ls2)[]{}; 
    \vertex [left=0.5cm of ls2](rms2)[]{$(\tau_2)$}; 
    \vertex [right=1.cm of ls2](v22)[sb]{\white\textbf\small $\mathbf{2}$}; 
    \vertex [right=1.8cm of ls2](v23)[sb]{\white\textbf\small $\mathbf{3}$}; 
    \vertex [right=3.2cm of ls2](v25)[sb]{\white\textbf\small $\mathbf{5}$}; 
    \vertex [right=5.0cm of ls2](rs2){}; 
    \vertex [above=0.8cm of l1](ls3)[]{}; 
     \vertex [left=0.5cm of ls3](rms3)[]{$(\tau_3)$}; 
     \vertex [right=2.5cm of ls3](v34)[sb]{\white\textbf\small $\mathbf{4}$}; 
     \vertex [right=5.0cm of ls3](rs3){}; 
    \vertex [left=0.5cm of ls](rms)[]{$(\eta\tau_1)$}; 
    \vertex [right=5.0cm of ls](rs){}; 
    \vertex [right=4.5cm of ls](vphi)[myblob2]{\white\textbf\small $\mathbf{8}$}; 
    \vertex [right=5.0cm of l1](r1)[]{}; 
    \vertex [right=0.5cm of ls](v1)[sb]{\white\textbf\small $\mathbf{1}$}; 
    \vertex [right=3.8cm of ls](v4)[sb]{\white\textbf\small $\mathbf{6}$}; 
   	 \diagram*{
   	 (ls)--[thick](v1)--[thick](v4)--[thick](vphi)--[thick](rs),(ls2)--[thick](v22)--[thick](v23)--[thick](v25)--[thick](rs2),(ls3)--[thick](v34)--[thick](rs3)};
    \end{feynman}  
  \end{tikzpicture} 
\end{equation*}
Reading the arguments of the {\it H} functions (by following the indices of the various horizontal lines), we arrive at the contributions
\begin{align}
  \langle\!\langle T_{(16),(235),(4)} \rangle\!\rangle &=
  \bar v\dd {H_{16}\over p^2_{81}-m^2}\, 
\dd {H_{235}(p_1,p_{68})\over p^2_{816}-m^2}\dd {H_{4}(p_{123},p_{568})\over p^2_{816235}-m^2}\dd u \,,
\end{align}
where the $H_{235}$ function reads:  
\begin{align}
    H_{235}(p_1,p_{68})&=2(p_{1}\mdot F_{235}\mdot p_{68})+(p_{1}\mdot F_{25})\dd (p_{2}\mdot F_{3})\nn\\
    &+(p_{1}\mdot F_{235})\dd p_{235}
+(p_{1}\mdot F_{23})\dd (p_{23}\mdot F_{5})+(p_{1}\mdot F_{2})\dd (p_{2}\mdot F_{35}) \, ,
\end{align}
as well as the $H_{16}$ function,
\begin{align}
    H_{16}= \big(p_8\mdot F_{16}\big) + {1\over 4}F_1\dd \big(p_1 \mdot F_{6}\big) \, ,
\end{align}
and $H_{4}$,
\begin{align}
   H_{4}(p_{123},p_{568}) =2 \big(p_{123}\mdot F_{4}\mdot p_{568}\big)+
 \big(p_{123}\mdot  F_{4}\big)\dd p_{4} \, .
\end{align}\\[5pt]
We conclude this section by presenting the following compact expression for numerators based on the above rules for evaluation: 
\begin{align}\label{eq:recJ}
   & \npre_{\!f}(12\ldots n{-}2, \sc {n\!-\!1}, \sc {n})=(-1)^{n\!-\!3} \ \bar v \dd\Big[ { H_{1\ldots n\!-\!2}\over p_{n1}^2-m^2} \Big]\dd u \nn\\
   &+\sum_{\tau_R\subset \{2,\ldots,n\!-\!2\}}(-1)^{|\tau_R|} \ \bar v \dd \Big[
   {{\cal X}(1\rho(\tau_R))\dd  H_{\tau_R}(p_{\Theta_L(\tau_R)},p_{\Theta_R(\tau_R)})\over p^2_{n1\tau_L}-m^2}\Big]\dd u \,.
\end{align} 
We have explicitly checked the compact numerator expressions provided by this formula up to eight points. We also note that in all cases the fermion color-kinematic numerator tends to the scalar Yang-Mills case in \cite{Chen:2022nei} after the replacement $\bar v\gamma^\mu u\to p_n^\mu$ and $\bar v\gamma^{\mu_1}\gamma^{\mu_2}\cdots \gamma^{\mu_r} u\to 0$ as expected for consistency.
\section{Conclusion and Outlook}
Using a universal method, we have presented a new and compact construction for a color-kinematic numerator associated with the generalized massive Compton amplitude with a fermion pair. We have verified evaluation maps explicitly until eight points and also presented a more straightforward graphical method for computing such numerators.

As the formulation relies on the Lorentz group representations of the matter fields, one can apply similar techniques to miscellaneous Compton scattering amplitudes, for instance, those with two massive vectors or other massive particles. An exciting track is to investigate if it is possible to shed new light on gauge and gravity interactions in high-spin matter fields at finite spin order. Studying the classical limit for such amplitudes and resolving questions related to consistent factorization limits could potentially aid the development of more efficient computational methods for observing spinning black hole mergers. It is beyond this paper's scope, so we leave these and related questions for future studies.

\section{Acknowledgements}
The work of N.E.J.B.-B. and G.C. was supported by the DFF grant 1026-00077B and partially by the Carlsberg Foundation. G.C. has received funding from the Marie Sklodowska-Curie grant agreement No. 847523 "INTERACTIONS" under the European Union Horizon 2020 research and innovation program.


\bibliographystyle{JHEP}
\bibliography{JHEPFinal/KinematicAlgebraFinal}

\end{document}